\documentclass[fleqn,usenatbib]{mnras}
\usepackage[modulo]{lineno}
\usepackage{ulem}
\usepackage{soul,xcolor}
\usepackage{xcolor}
\usepackage[T1]{fontenc}
\usepackage{graphicx}	
\usepackage{amsmath}	
\usepackage{amssymb}	
\usepackage{longtable}
\DeclareRobustCommand{\VAN}[3]{#2}
\let\VANthebibliography\thebibliography
\def\thebibliography{\DeclareRobustCommand{\VAN}[3]{##3}\VANthebibliography}

\newcommand{\teff}{$T_{\rm eff}$} 
\newcommand{\logg}{$\log g$}

\newcommand{\metal}{\mbox{[Fe/H]}}
\newcommand{\cfe}{\mbox{[C/Fe]}}

\title[2MASS~J1808$-$5104 the most primitive thin disk star]{{The chemical abundance pattern of the extremely metal-poor thin disk star 2MASS~J1808$-$5104 and its origins}\thanks{This paper includes data gathered with the 6.5\,m Magellan Telescopes located at Las Campanas Observatory, Chile.}}
\author[Mohammad. K.\ Mardini et al.]{
Mohammad K.\ Mardini,$^{1,2}$\thanks{E-mail: m.mardini@ipmu.jp}
Anna Frebel,$^{3}$
Rana Ezzeddine,$^{4}$
Anirudh Chiti,$^{5,6}$
Yohai Meiron,$^{7}$
\newauthor
~Alexander~P.~Ji,$^{5}$
Vinicius M.\ Placco,$^{8}$
Ian U. Roederer,$^{9,10}$
and Jorge Meléndez$^{11}$
\\
$^{1}$Kavli IPMU (WPI), UTIAS, The University of Tokyo, Kashiwa, Chiba 277-8583, Japan\\
$^{2}$Institute for AI and Beyond, The University of Tokyo 7-3-1 Hongo, Bunkyo-ku, Tokyo 113-8655, Japan\\
$^{3}$Department of Physics and Kavli Institute for Astrophysics and Space Research, Massachusetts Institute of Technology, Cambridge, MA 02139, USA\\
$^{4}$Department of Astronomy, University of Florida, Bryant Space Science Center, Gainesville, FL 32611, USA\\
$^{5}$Department of Astronomy $\&$ Astrophysics, University of Chicago, 5640 S Ellis Avenue, Chicago, IL 60637, USA\\
$^{6}$Kavli Institute for Cosmological Physics, University of Chicago, Chicago, IL 60637, USA\\
$^{7}$SciNet High Performance Computing Consortium, University of Toronto, 661 University Ave., Toronto, ON M5G\,1M1, Canada\\
$^{8}$NSF’s NOIRLab, 950 N. Cherry Ave., Tucson, AZ 85719, USA\\
$^{9}$Department of Astronomy, University of Michigan, 1085 S. University Ave., Ann Arbor, MI 48109, USA\\
$^{10}$Joint Institute for Nuclear Astrophysics -- Center for the Evolution of the Elements (JINA-CEE), USA\\
$^{11}$Instituto de Astronomia, Geofísica e Ciências Atmosféricas, Universidade de São Paulo, 05508-090, São Paulo, Brazil
}

\pubyear{2015}

\begin{document}
\label{firstpage}
\pagerange{\pageref{firstpage}--\pageref{lastpage}}
\maketitle

\begin{abstract}
We present a high-resolution ($R\sim35,000$), high signal-to-noise
($S/N=350$) Magellan/MIKE spectrum of the bright extremely metal-poor star
2MASS~J1808$-$5104. We find [Fe/H] = $-$4.01 (spectroscopic LTE stellar parameters), [Fe/H] = $-$3.8 (photometric stellar parameters), [Fe/H] = $-$3.7 (spectroscopic NLTE stellar parameters). We measured a carbon-to-iron ratio of $\mbox{[C/Fe]}= 0.38$ from the CH G-band. J1808$-$5104 is thus not carbon-enhanced, contrary to many other stars with similarly low iron abundances. We also determine, for the first time, a barium abundance ($\mbox{[Ba/Fe]} =-0.78$), and obtain a significantly reduced upper limit for the nitrogen abundance ([N/Fe]$ < - 0.2$). J1808$-$5104 has low ratio of $\mbox{[Sr/Ba]}=-0.17$, which is consistent with that of stars in ultra-faint dwarf galaxies. We also fit
the abundance pattern of J1808$-$5104 with nucleosynthesis yields from a grid of Population\,III supernova models. There is a good fit to the abundance pattern which suggests J1808$-$5104 originated from gas enriched by a single massive supernova with a high explosion energy of E $=10\times10^{51}$\,erg and a progenitor stellar mass of M$=29.5$\,M$_{\odot}$. 
Interestingly, J1808$-$5104 is a member of the Galactic thin disk, as confirmed by our detailed kinematic analysis and calculated stellar actions and velocities. Finally, we also established the orbital history of J1808$-$5104 using our time-dependent Galactic potential the \texttt{ORIENT}. J1808$-$5104 appears to have a stable quasi-circular orbit and been largely confined to the thin disk. This unique orbital history, the star's very old age ($\sim13.5$\,Gyr), and the low [C/Fe] and [Sr/Ba] ratios suggest that J1808$-$5104 may have formed at the earliest epoch of the hierarchical assembly of the Milky Way, and it is most likely associated with the primordial thin disk.
\end{abstract}




\section{Introduction}
The chemical abundances of the most metal-poor stars trace the earliest nucleosynthesis events of
elements heavier than H and He, which took place within the first billion years after the Big 
Bang \citep{alvarez06,Becker2012}. Stars with ${\metal} \sim -4.0$ and less (also known as Ultra Metal-Poor - UMP) are best suited for this, as they are likely second-generation stars, thus enabling the study of their massive and short-lived progenitor Population\,III (Pop\,III)
stars \citep{fn15}. 
Measuring as many chemical elements as possible in these stars thus helps to constrain models of Pop\,III nucleosynthesis. Carbon, in particular, has played an important role in learning about progenitor properties. Large observed carbon abundances have been interpreted as a signature of mixing and fallback supernovae \citep[e.g.,][]{UmedaNomotoNature,heger10,cooke14}.
On the contrary, other types of supernovae must have been responsible for the abundance patterns observed in ultra-metal-poor stars that are not carbon enhanced \citep[e.g.,][]{Caffau2011,placco2021,skuladottir2021}. Their formation mechanism might also have been entirely different, and e.g., not driven by carbon and oxygen-cooled gas \citep{brommnature, dtrans} but through dust cooling \citep{chiaki15,debennassuti14,ji14}.

Indeed, for the ${\metal} \lesssim -4.0$ metallicity regime, about 81\% of the $\sim40$ stars observed to date are carbon enhanced, i.e., have \cfe$>0.7$ \citep{placco14,arentsen2022}. Restricting stars to have halo kinematics increases the carbon fraction to $\sim 90\%$. This difference can be attributed to a handful of stars being in fact associated with the metal-poor Atari disk population \citep[for more details, see][]{Mardini2022}. Interestingly, these stars are all non-carbon-enhanced. One star is associated with the thin disk, which is the subject of this paper -- it is also not carbon-enhanced (see Section~\ref{sec:analysis}). 

\citet{melendez16} reported the discovery of 2MASS~J18082002$-$5104378 (hereafter J1808$-$5104), with ${\metal} =-4.07$ and upper limits on the carbon abundance of $\mbox{[C/Fe]}<0.94$ and the barium abundance $\mbox{[Ba/Fe]}<-0.31$, based on an ESO/UVES high-resolution spectrum. Their upper limit on the carbon abundance indicated this star might fall into the category of stars
without carbon-enhancement, but firm conclusions could not be drawn. More recently, \citet{Spite2019} confirmed a mild enhancement of J1808$-$5104 of $\mbox{[C/Fe]}=0.49$ and also reported oxygen and beryllium measurements. Both studies were able to detect strontium (Sr\,{II}, Z=38), but no other neutron-capture elements were reported. 

Here, we report on results of our new high-resolution, high signal-to-noise spectroscopic observations with the Magellan telescope, which confirmed the mild carbon enhancement and enabled a barium (Ba\,{II}, Z=56) detection for the first time. A low upper limit of zinc (Zn\,{I}, Z=30) is also found which is unusual at this low metallicity. J1808$-$5104 has a low barium abundance and is thus not a neutron-capture element enhanced star. Our observations also produced upper limits on additional neutron-capture elements yttrium (Y\,{II}, Z=39) and europium (Eu\,{II}, Z=63), as well as nitrogen which all help constrain the main characteristics of the Pop\,III stellar progenitors \citep{placco16}. 

Dynamically, and based on Z$_{max}$ values, \citet{Schlaufman2018} speculated that J1808$-$5104 is confined within the thin disk. The disk-like kinematics were later confirmed by \citet{Sestito2019}. Also, \citet{Schlaufman2018} confirmed the binarity of J1808$-$5104 by investigating its radial velocity using 14 Magellan/MIKE, the three VLT/UVES spectra observed by \citet{melendez16}, and 31 GMOS/Gemini South spectra. \citet{Spite2019} provided additional radial velocity measurements for the UVES R760 spectrum, and confirm velocity variations for J1808$-$5104.
Chronologically, \citet{Schlaufman2018} used isochrones to estimate an age of 13.5\,Gyr for J1808$-$5104. 

In this paper, we report on a further abundance analysis. Details of our spectroscopic observations are provided in Section~\ref{sec:obs}, stellar parameters and chemical abundances are described in Sections~\ref{sec:chem} and \ref{sec:discussion}, respectively, and our conclusion that J1808$-$5104 is an extremely-metal-poor star showing an abundance signature typical for metal-poor stars formed from well-mixed gas and belong to the thin disk is presented in Section~\ref{sec:Conclusions}.

\section{Observations and Radial Velocity Measurements}\label{sec:obs}

\begin{figure}
\begin{center}
\includegraphics[clip=true,width=7.5cm]{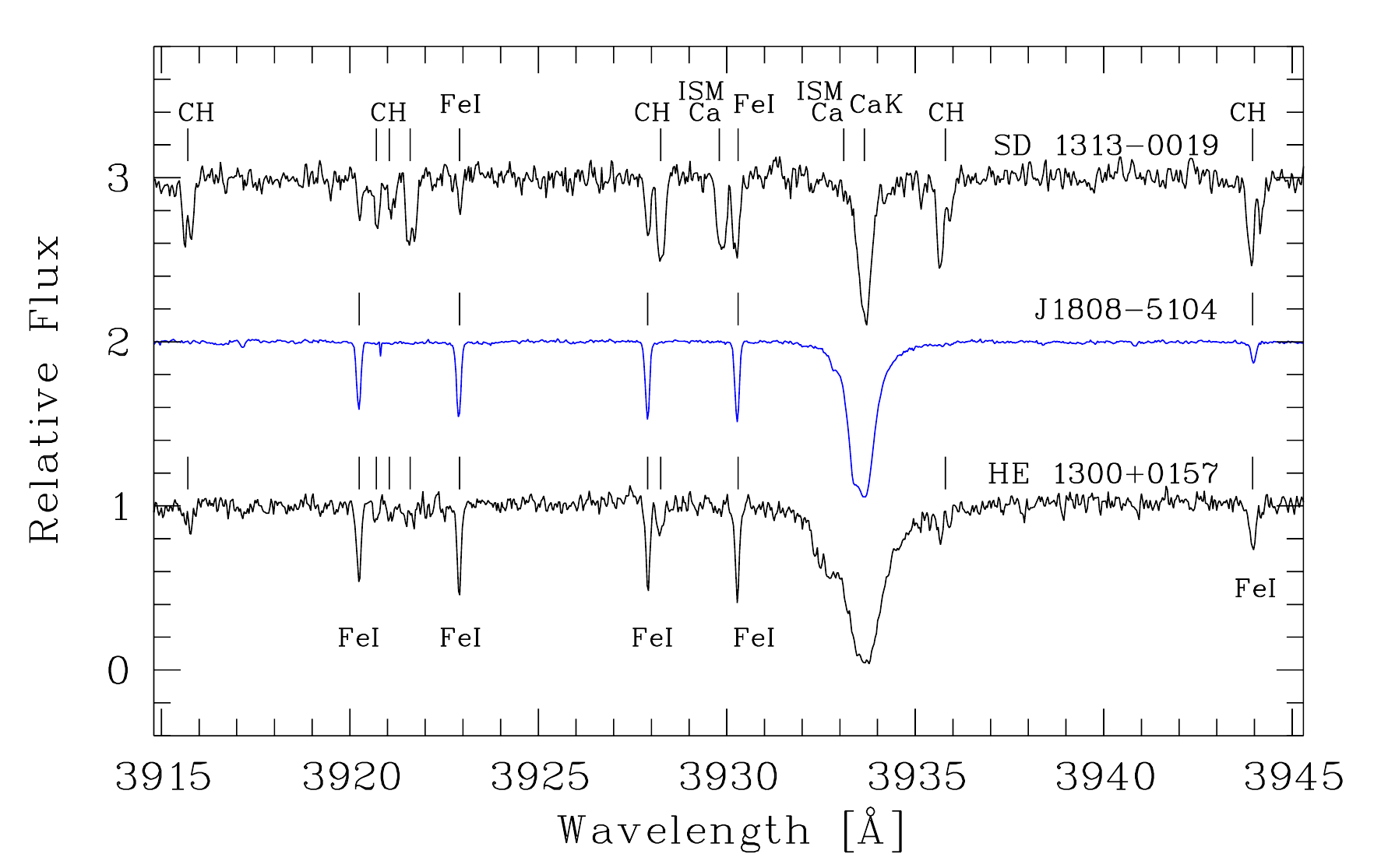}\\ 
\includegraphics[clip=true,width=7.5cm]{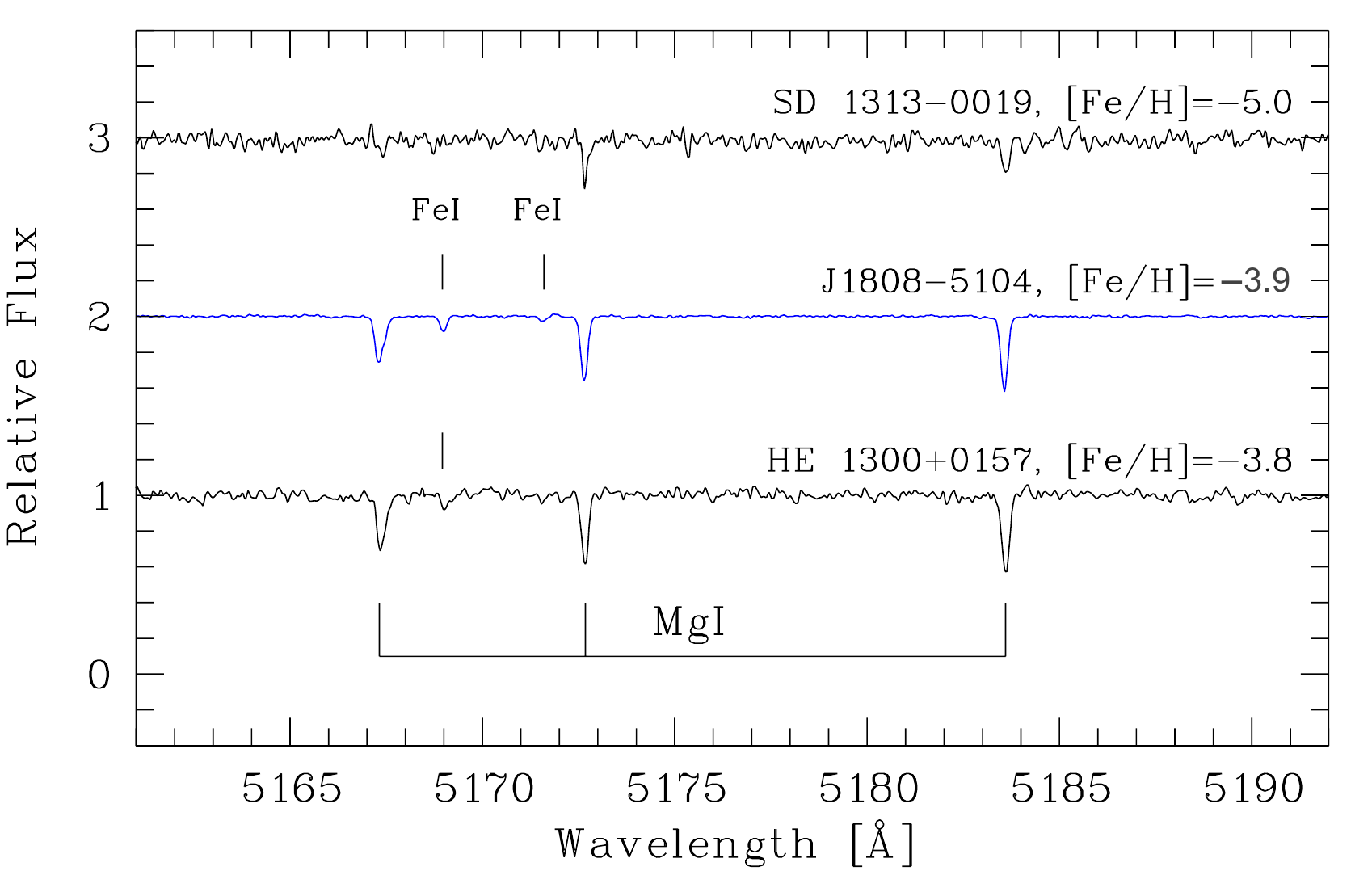}\\ 
\includegraphics[clip=true,width=7.5cm]{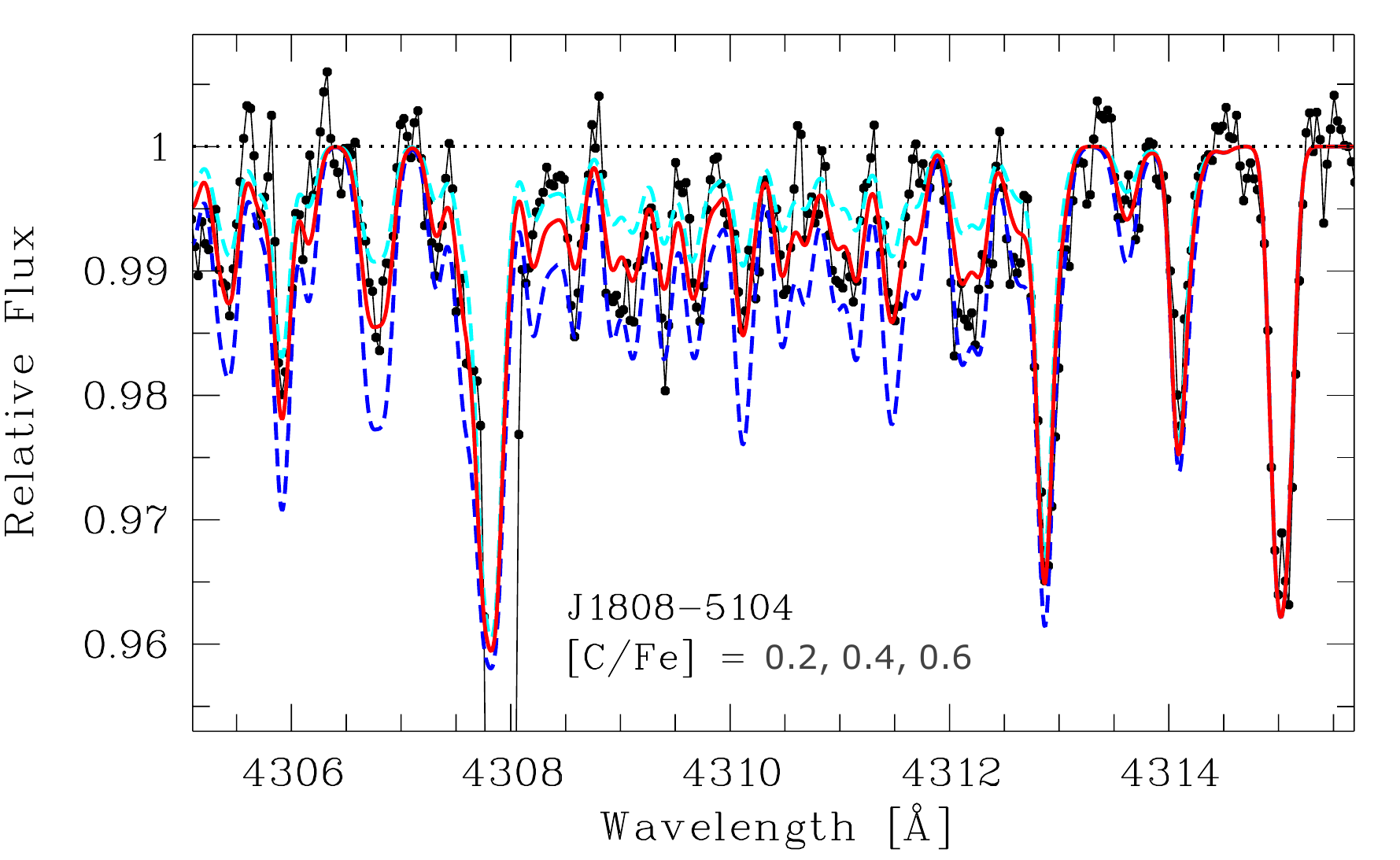}\\
\includegraphics[clip=true,width=7.5cm]{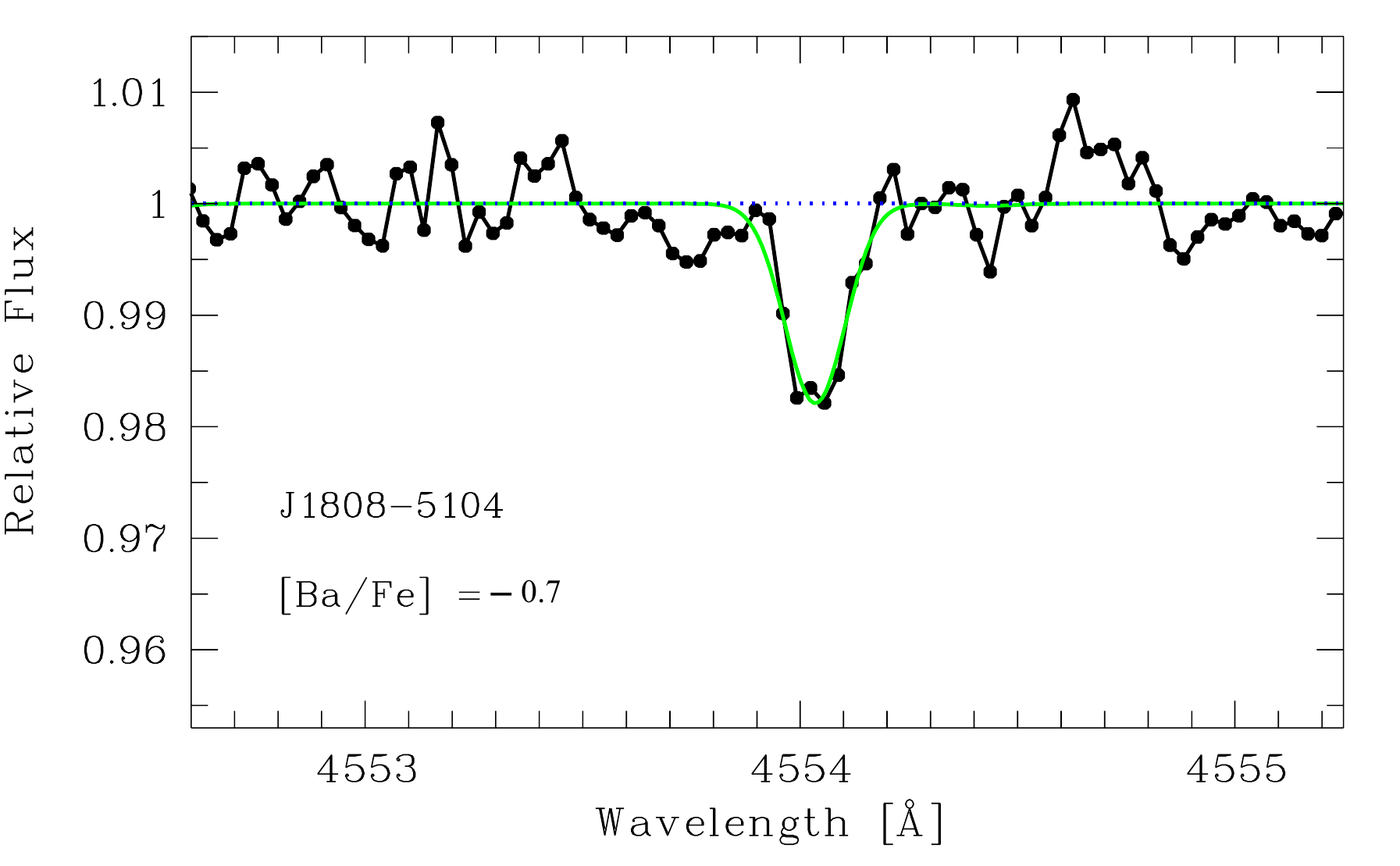}
\caption{Portions of the Magellan/MIKE spectrum of
J1808$-$5104 in comparison with other iron-poor stars near the
Ca\,II\,K line at 3933\,{\AA} (top) and around the Mg\,b lines at
5180\,{\AA} (middle panel). Also, we show the spectral syntheses used for the determination of carbon and barium abundances using the G-band near 4313\,{\AA} (middle panel), and the Ba\,II line at 4554\,{\AA} (bottom), respectively. \label{specs}}
\end{center}
\end{figure}

We observed J1808$-$5104 (R.A. = 18:08:20.02, Dec. = $-$51:04:37.8,
$V=11.9$) with the MIKE spectrograph on the Magellan-Clay
telescope at Las Campanas Observatory on April 15, 16, and 17, 2016,
for a total of 2\,h during clear weather and seeing conditions varying
from 0\farcs5 to 0\farcs7. The employed $0\farcs7$ slit yields a high
spectral resolution of $\sim30,000$ in the red and $\sim35,000$ in the
blue wavelength regime of our spectrum, covering 3300\,{\AA} to
9400\,{\AA}. Data reductions were carried out with the MIKE Carnegie
Python pipeline \citep{kelson03}\footnote{Available at \url{http://obs.carnegiescience.edu/Code/python}}. The resulting $S/N$
per pixel is $\sim65$ at $\sim3500$\,{\AA}, $\sim200$ at
$\sim4000$\,{\AA}, $\sim350$ at $\sim4700$\,{\AA}, $\sim270$ at
$\sim5200$\,{\AA}, and $\sim420$ at $\sim6000$\,{\AA}. In
Figure~\ref{specs} we show several representative portions of the
J1808$-$5104 spectrum, around the Ca\,{II}\,K line at 3933\,{\AA}, the Mg\,b lines at
5170\,{\AA}, the G-bandhead at 4313\,{\AA}, and the Ba line at
4554\,{\AA}, compared with the spectra of SD~1313$-$0019 \citep{frebel15b} and 
HE~1300$+$0157 \citep{frebel07}.


\begin{figure}
\begin{center}
\includegraphics[width=0.5\textwidth]{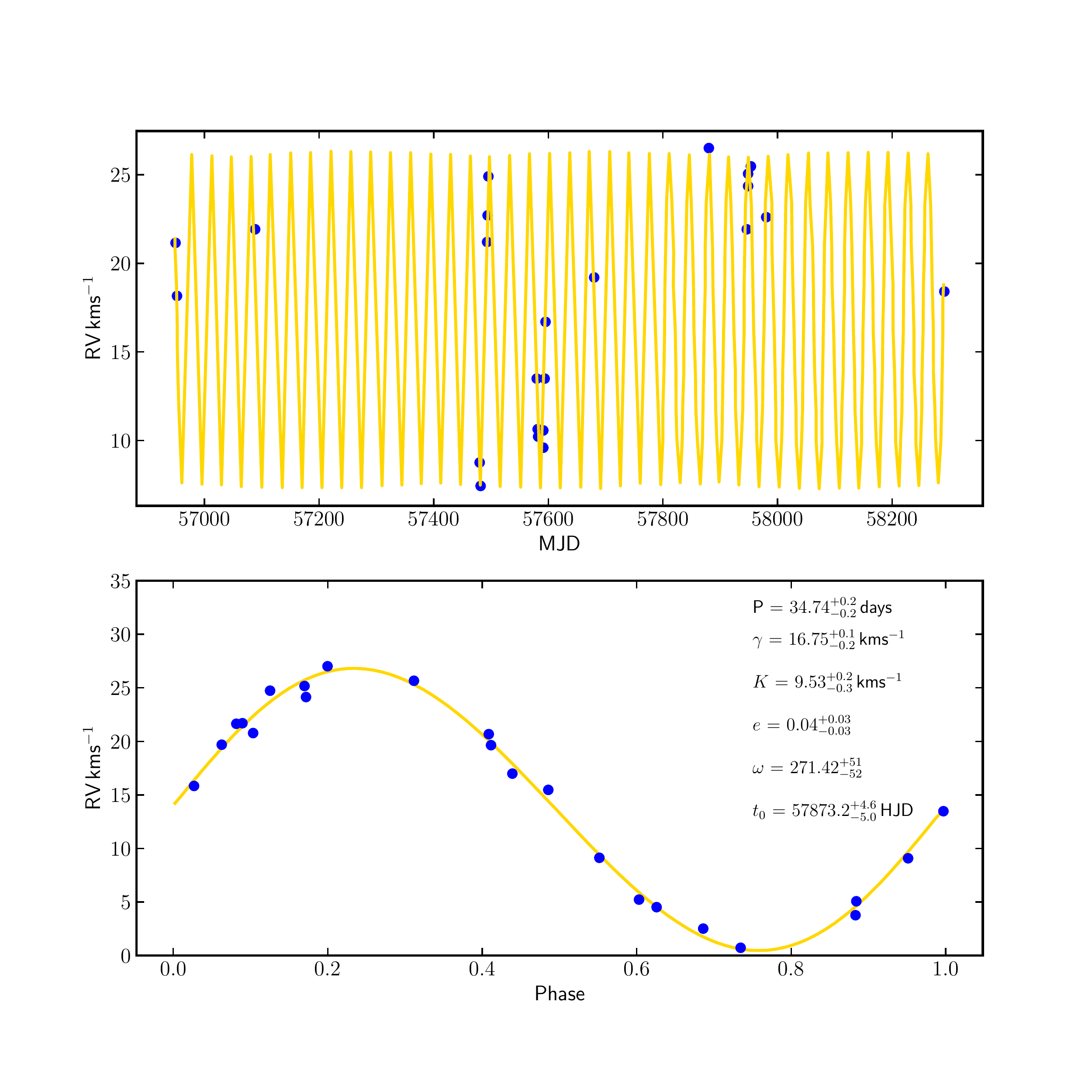}
\caption{Keplerian orbit for the 2MASS J1808$-$5104 system using radial velocities measurements reported by \citet{Schlaufman2018}, \citet{Spite2019}, and this work. Gold lines represent the best fit. Blue points represent the radial velocity measurements. Orbital parameters are reported in the lower panel.}
\label{orbital_period}
\end{center}
\end{figure}

We measured the radial velocity in our three individual spectra taken on consecutive nights, and confirm
velocity variations reported by \citet{Schlaufman2018} and \citet{Spite2019}. Our heliocentric values are 21.2\,km\,s$^{-1}$
(2016 April 15), 22.7\,km\,s$^{-1}$ (2016 April 16), and
24.9\,km\,s$^{-1}$ (2016 April 17). We furthermore obtained followup
observations to make additional radial velocity measurements. We
find 26.5\,km\,s$^{-1}$ (2017 May 7) and 22.6\,km\,s$^{-1}$ (2017 Aug
15). We fit a Keplerian orbit to all available radial velocities measurements from the literature, based on high resolution spectra (see table~2 in \citealt{Schlaufman2018} and \citealt{Spite2019}) using the program \texttt{BinaryStarSolver} \citep{Milson2020}. Figure~\ref{orbital_period} shows the fitted Keplerian orbit overplotted with all the data.

We find an orbital period of $P = 34.7385_{-0.2}^{+0.2}$\,days, a system velocity $\gamma = 16.745_{-0.2}^{+0.1}$\,km s$^{-1}$, a velocity semi-amplitude $K = 9.53_{-0.2}^{+0.3}$\,km s$^{-1}$, an eccentricity $e = 0.039_{-0.03}^{+0.03}$, a longitude
of periastron $\omega = 271.42_{-52}^{+51}$ deg, and a time of periastron
$t_{0} = 57873.2_{-4.9}^{+4.6}$ HJD. We also calculated the projected
semimajor axis $a_{1} \sin{i} = 4.55_{-0.2}^{+0.2}$\,R$_{\odot}$ and the mass function $f(M) = 0.0031_{-0.00034}^{+0.00034}$\,M$_{\odot}$. In general, these orbital parameters are in good agreement with the ones reported by \citet{Schlaufman2018} and \citet{Spite2019}. Thus, J1808$-$5104 the oldest thin-disk confirmed binary at the lowest metallicities. Overall, the number of binaries among metal-poor stars with ${\metal} \lesssim -2.0$ is close to 20\% \citep{hansen16}.

\section{Chemical abundance analysis}\label{sec:chem}
\subsection{Stellar parameters}\label{sec:stellpar}

We measured equivalent widths to obtain atmospheric chemical abundances, by fitting Gaussian profiles to absorption features. Results are listed
in Table~\ref{Tab:Eqw}. To perform the abundance determination, we used a 1D plane-parallel
model atmosphere with $\alpha$-enhancement \citep{castelli_kurucz} and the latest version of MOOG\footnote{\url{http://www.as.utexas.edu/~chris/moog.html}} \citep{moog, sobeck11}. Abundances of blended features were
determined with spectrum synthesis, using the Spectroscopy Made Hard (SMH) software package \citep{casey14}. The line lists for the atomic and molecular features were generated by the \texttt{linemake}\footnote{\url{https://github.com/vmplacco/linemake}} code \citep{placco2021b}. The isochrone fitting done by \citet{Schlaufman2018} suggested a slightly warmer {\teff} and more metal-rich {\metal} star than what was reported by \citet{melendez16}. \citet{Spite2019} also estimated their stellar parameters using Gaia\,DR2 photometry \citep{Gaia_dr2} and 3D maps of interstellar reddening \citep{Lallement2018} to be warmer and more metal-rich. We then employed three different methods to obtain
stellar parameters from Fe\,{I} and Fe\,{II} lines: (1) The commonly used technique of calculating line formation under the assumption of local thermodynamic
equilibrium (LTE). (2) The quantum fitting method (QFM) that invokes
non-local thermodynamic equilibrium (NLTE). (3) The procedure outlined in  \citet{roederer2018}.

\begin{table}
\centering
\caption{Line list and equivalent width measurements.}
\label{Tab:Eqw}
\begin{tabular}{lrrrrr}
\hline
Species & $\lambda$ &$\chi$ &$ \log gf$ & EW    & $\log\epsilon$ (X) \\
        &	(\AA)   & (eV)  &    (dex)  &(m\AA) & (dex) \\
\hline
Li\,I&  6707.7  &  0.00    & 0.170     & syn       &      1.35 \\
CH  &   4313    &   \ldots &  \ldots   & syn       &      5.00 \\
CH  &   4323    &   \ldots &  \ldots   & syn       &      4.95 \\
NH  &   3360    &   \ldots &  \ldots   & syn       &    $<$3.67 \\ 
O\,I&    7771.94&   9.15&   0.37& syn     & 6.00: \\
O\,I&    7774.17&   9.15&   0.22& syn     & 6.20: \\
Na\,I&   5889.95&   0.00&   0.11&    39.73& 2.12 \\
Na\,I&   5895.92&   0.00&$-$0.19&    21.15& 2.04 \\
Mg\,I&   4057.51&   4.35&$-$0.90&    2.45 & 4.09 \\
Mg\,I&   4167.27&   4.35&$-$0.74&    3.71 & 4.10 \\
Mg\,I&   4702.99&   4.33&$-$0.44&    6.89 & 4.04 \\
Mg\,I&   5172.68&   2.71&$-$0.36&    81.18& 4.14 \\
Mg\,I&   5183.60&   2.72&$-$0.17&    89.43& 4.10 \\
Mg\,I&   5528.40&   4.35&$-$0.55&    6.26 & 4.10 \\
Mg\,I&   8806.76&   4.35&$-$0.14&    18.88& 4.10 \\
Al\,I&   3944.00&   0.00&$-$0.64&    18.32& 1.62 \\
Al\,I&   3961.52&   0.01&$-$0.33&    28.60& 1.58 \\
Si\,I&   3905.52&   1.91&$-$1.04& syn     & 3.72 \\
Si\,I&   4102.94&   1.91&$-$3.34& 0.9     & 3.89 \\
\hline
\end{tabular}
\\
This table is published in its entirety in the electronic edition of the paper. A portion is shown here for guidance regarding its form and content.
\end{table}

\subsubsection{LTE Stellar Parameters}\label{lte_stellpar}

We initially derived the stellar parameters ({\teff} = 5070\,K, {\logg} = 2.40\,dex, {\metal} = $-4.20$\,dex, and $v_{\text{micro}}$ = 1.30\,km s$^{-1}$) by (1) minimizing the trend between the reduced equivalent width and excitation potential for the Fe\,{I} lines, and (2) forcing agreement between Fe\,{I} and Fe\,{II} abundances. The initial {\teff} was then adjusted following the photometric correction presented in \citet{Frebel2013}. The rest of the stellar parameters were adjusted to produce no trend of Fe\,{I} abundances with reduced equivalent width. We also adjusted the Fe\,{II} abundance to match again the Fe\,{I} abundance. This yields {\teff} = 5233\,K, {\logg} = 2.80\,dex, {\metal} = $-4.01$\,dex, and $v_{\text{micro}}$ = 1.35\,km s$^{-1}$.

\subsubsection{NLTE Stellar Parameters}\label{nlte_stellpar}

We use the QFM method developed by \citet{ezzeddine16} for the iron abundance to spectroscopically determine a second set of stellar parameters. The Fe\,{I}/Fe\,{II}/Fe\,{III} model atom used in \citet{ezzeddine16} is compiled from a large number of energy levels taken from the NIST\footnote{\url{http://www.nist.gov/}} database (846 Fe\,{I}, 1027 Fe\,{II}
and the Fe\,{III} continuum) and theoretical levels from
\citet{petkur15}, and reduced into a super-atom of 424 total
levels. Levels are coupled radiatively via a large number of
bound-bound transitions ($\sim$25,000 lines from the VALD3
database\footnote{\url{http://vald.astro.uu.se/}}) and photoionization
tables. Additionally, all levels are coupled collisionally via
inelastic electron and hydrogen collisions. Hydrogen collisional rates
are estimated using the new semi-empirical QFM. It includes the
dominating ion-pair production process, which avoids the large
uncertainties usually obtained when using the classical
\citet{drawin1968,drawin1969a,drawin1969b} approximation. We refer the
interested reader to \citet{ezzeddine16} for a more detailed
description of the atom.

Departures from LTE for each individual Fe\,{I} and Fe\,{II} line were
calculated to iteratively determine the stellar parameters
spectroscopically, just as in the LTE case. We follow the procedure
described in \citet{ezzeddine17} who applied it to the 20 most
iron-poor stars. As for these stars, the scatter among line abundances of J1808$-$5104 is reduced compared to the LTE analysis. The standard deviation of Fe\,{I} line abundances is 0.05\,dex for the NLTE analysis, compared to the already low $0.07$\,dex LTE result. This further supports applying quantum mechanically based NLTE corrections to individual lines, as it leads to improved overall results, independent of the data quality. 

Based on the NLTE Fe\,{I} line analysis, we spectroscopically obtain T$_{\rm
  eff}=5250$\,K.  This yields an Fe\,{I} abundance of
${\metal}=-3.69$, and $-3.65$ for Fe\,{II}. The NLTE Fe\,{I} abundance is higher by 0.46\,dex compared to the Fe\,{I} LTE case. Fe\,{II} lines are hardly affected by NLTE, at the level of 0.02\,dex which we neglect. This difference (i.e., $\Delta\mbox{[FeI/H]} = 0.46$) is in line with what can be obtained from the relation of $\Delta\mbox{[FeI/H]} = -0.14 \times
{\metal}_{\rm{LTE}} - 0.15$ developed by \citet{ezzeddine16}, $\Delta{\metal} =0.43$. It illustrates the strong metallicity dependence of departures from LTE. As a consequence of the differential NLTE effect for Fe\,{I} and Fe\,{II}, the surface gravity is somewhat higher, $\log g=3.2\pm0.2$ since our spectroscopic analysis aims for an ionization balance of Fe\,{I} and II in NLTE. The microturbulence is somewhat higher also, $v_{micro}=1.8\pm0.2$\,km\,s$^{-1}$.

\subsubsection{Photometric Stellar Parameters}\label{rpa_stellpar}

Photometric stellar parameters were also obtained for J1808$-$5104 based on a procedure described in detail by \citet{roederer2018}. In summary, the effective temperature (\teff) was calculated with the color--\metal--\teff\ relations of \citet{casagrande2010}, using the $JHK$ magnitudes from 2MASS \citep{cutri2003}, the Johnson $BV$ magnitudes from APASS DR9 \citep{henden2014}, and the corrected \metal\ value from Section~\ref{lte_stellpar} as a first-pass estimate. The \logg\ was calculated from the fundamental relation described in \citet{roederer2018}, using the 3D reddening values, $E(B-V)$, from the \texttt{bayestar2017} version of the \texttt{\href{https://dustmaps.readthedocs.io/en/latest/}{dustmaps}} application \citep{green2018}. The distance was taken from \citet{bailer-jones2021} and the bolometric correction in the $V$ band (BC$_V$) from \citet{casagrande2014}. These initial \teff\ and \logg\ values are used to derive the \metal\ and microturbulent velocity ($v_{\text{micro}}$). Then, the first-pass \metal\ estimate is updated and \teff\ and \logg\ are recalculated. This yields {\teff} = 5665\,K, {\logg} = 3.34\,dex, {\metal} = $-3.85$\,dex, and $v_{\text{micro}}$ = 1.52\,km s$^{-1}$. Since the \teff\, and \logg\, are determined independently from the model atmospheres, we have decided to adopt the parameters above for the remainder of the analysis presented in this paper.

We estimate random uncertainties for the stellar parameters by varying only one parameter at a time until $1\sigma$ scatter in the previous procedure is achieved. In general, our stellar parameters agree well with those of \citet{Spite2019}, who derive $\sim5600$\,K $\log g=3.4$, $v_{\text{micro}}=1.6\pm0.2$\,km\,s$^{-1}$, and ${\metal}=-3.84\pm0.07$.

\subsection{Chemical abundances}\label{sec:analysis}

\begin{figure*}
\begin{center}
\includegraphics[clip=true,width=18cm]{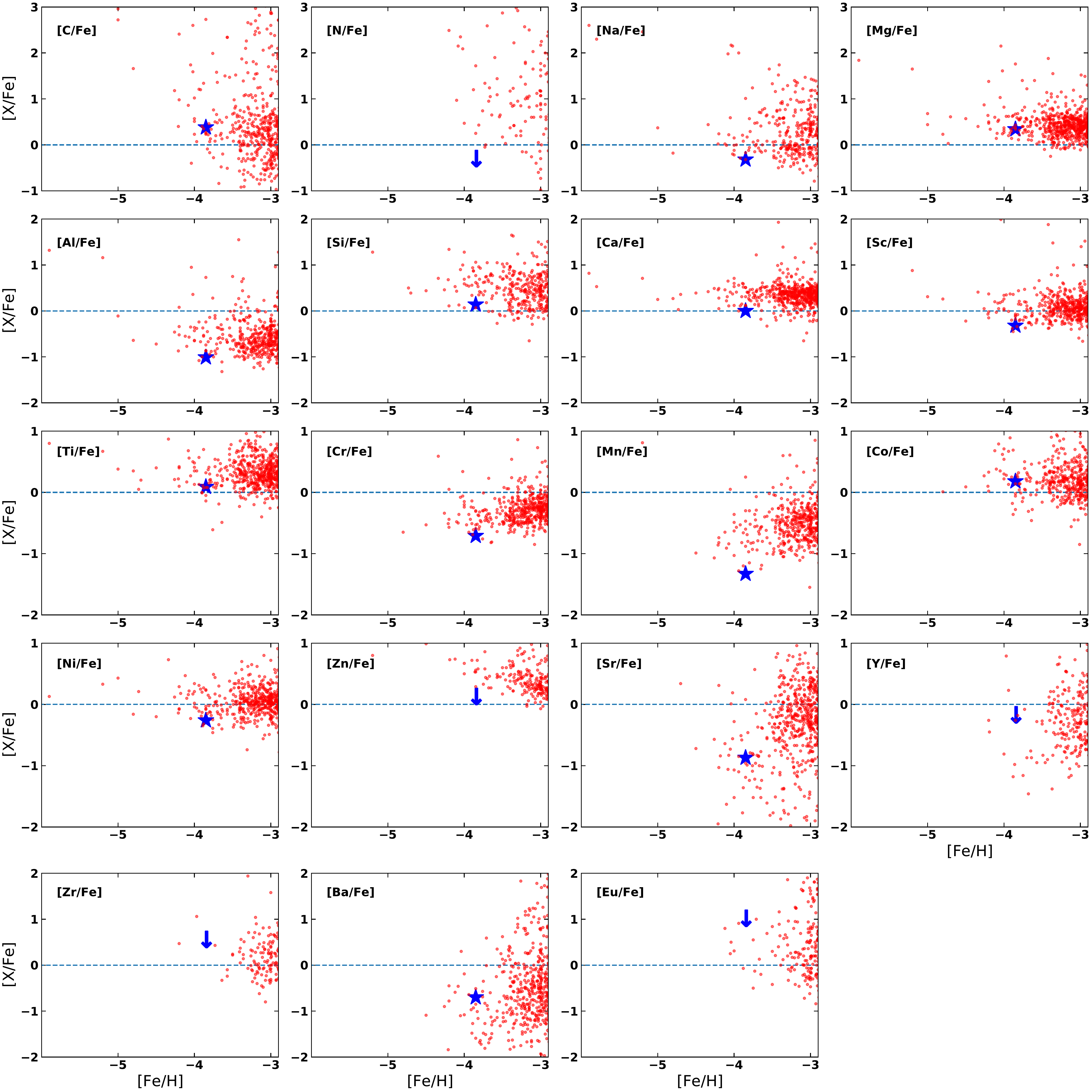} 
\caption{Abundance ratios (\mbox{[X/Fe]}) as a function of metallicity ({\metal}) for various elements detected in J1808$-$5104 (blue asterisk) and other metal-poor stars collected by JINAbase \citep{jinabase}. Every red point represents a star taken from: \citet{roederer14c}; \citet{barklem05}; \citet{Ryan1991}; \citet{Cohen2004}; \citet{Hollek2011}; \citet{Bonifacio2012}; \citet{Mardini_2019a}; \citet{Aguado2017b}; \citet{Mardini_2019b}; \citet{Yong2013}; \citet{Frebel2010}; \citet{Hansen2015}; \citet{Mardini_2020}; \citet{Jacobson2015};
\citet{ezzeddine19};
\citet{Spite2014}; \citet{cayrel2004}; \citet{Norris2007}; \citet{Mardini2019c}; \citet{Aguado2021}; \citet{Lai2004}; \citet{Almusleh2021}; \citet{Placco2020}; \citet{Ryan1996}; \citet{Aguado2018}; \citet{Chiti2022}; \citet{Taani2019,Taani2019b}; \citet{Aoki2007}; \citet{Aoki2013}; \citet{Cohen2013}; \citet{Honda2011}; \citet{Lai2008}; \citet{Casey2015}; \citet{Li2015a}; \citet{Masseron2006}; \citet{Rich2009}; \citet{Taani_2022}; \citet{Ryan1999}; \citet{Spite1999}; \citet{Depagne2000}; \citet{Spite2000}; \citet{Sivarani2006}; \citet{Norris2001}; \citet{Taani2016RAA}; \citet{Placco2014}; \citet{Caffau2011a}; \citet{frebeletal05}; \citet{frebel08}; \citet{Li2015}; \citet{Behara2010}; \citet{Caffau2011}; \citet{Taani2020}; \citet{Susmitha2016}; \citet{Aguado2017}; \citet{Carretta2002}; \citet{For2010}; \citet{2015PASJ...67...84L}; \citet{Keller2014}; \citet{frebel15b}; \citet{Caffau2013}; \citet{Plez2005}; \citet{Christlieb2002}; \citet{Frebel2019}.    \label{fig:abundplot}}
\end{center}
\end{figure*}

We determined chemical abundances of 19 elements and five upper limits
for J1808$-$5104. The final abundance ratios [X/Fe] are calculated
using the photometric stellar parameters and solar abundances of \citet{asplund09}, and listed in Table~\ref{Tab:abund}. They are also shown in Figure~\ref{fig:abundplot} where we compare them with literature data.  In the following, we comment on the various observed element abundances and measurement details.

J1808$-$5104 is a warm metal-poor star. In accordance with its stellar evolutionary status at the bottom of the giant branch, we find a
lower-than-Spite Plateau \citep{spite82} abundance of A(Li) = 1.38, as measured from
the lithium doublet at 6707\,{\AA}. For comparison, A(Li) = 1.5 and 1.78 were found by
\citet{melendez16} and \citet{Spite2019}, respectively.

Since our spectrum has a very high S/N ratio, a carbon abundance could be
clearly measured from the CH G-bandhead at 4313\,{\AA}, yielding
$\mbox{[C/Fe]}=0.38 \pm 0.10$. Our [C/Fe] is in agreement with the one reported by \citet{Spite2019} of $\mbox{[C/Fe]}=0.49$. The detection is shown in Figure~\ref{specs},
together with the best-fit synthetic spectrum. To achieve the best possible
detection, we note that we co-added the 2016 spectrum with the two 
radial-velocity spectra taken in 2017. Adding the extra data somewhat increased the S/N and aided the measurement. Given the relatively unevolved nature of J1808$-$5104, there is no need to correct the carbon abundance for the star's evolutionary status \citep{placco14} to obtain its true birth carbon abundance. As already noted in \citet{Spite2019}, 1D LTE abundances of
molecular species, such as CH, potentially suffer from strong effects
from not employing 3D model atmospheres. We thus note here as well that the 3D LTE abundance would be even lower. From Table~2 in \citet{gallagher16}, we estimate a potential correction of $-0.5$\,dex (by coarsely extrapolating their 5900\,K/4.0 model to our values). However, it remains to be seen what any 3D-NLTE abundances derived from CH might be, if available.

We obtained an upper limit for nitrogen from the non-detection of the NH band at 3360\,{\AA}. The still reasonable $S/N \sim 30$ yields a subsolar limit of [N/Fe]$<-0.2$ which shows J1808$-$5104 to be deficient in N. We also obtain a tentative oxygen abundance of $\mbox{[O/Fe]} = 1.25 \pm{0.50}$ from the two stronger of the near infrared O triplet lines at 7771 and 7774\,{\AA}, as they are very weakly detected only. This {$\mbox{[O/Fe]}$} value is in principally agreement with the one reported by \citet{Spite2019} of ${\mbox{[O/Fe]}}=1.36$ based on UV-OH line. However, we caution that 3D corrections would affect the OH abundance whereas NLTE significantly affects the triplet lines. While we do not further assess these corrections, we conservatively conclude that J1808$-$5104 is at least mildly oxygen enhanced, very similar to other metal-poor stars \citep{garciaperez_primas2006_O}.

There is some interstellar absorption by Na and Ca, as seen in the spectrum of J1808$-$5104. Both stellar Na D lines are very close to the strong interstellar lines but their equivalent widths could still be measured. Figure~\ref{specs} shows that the Ca\,II K line appears somewhat distorted by an interstellar component. However, we measure the Ca abundance from seven Ca\,I lines, so the blending of the Ca\,II\,K line is a purely cosmetic effect. As for abundances of other light elements, lines are generally weak but measurable given the good $S/N$. Seven Mg lines across the spectrum could be used to determine the Mg abundance. Given the low C abundance, the Al line at 3944\,{\AA}, which is blended with a CH feature, could be easily used in addition to the $\lambda$3961\,{\AA} line. For our star, the derived Al abundances for both lines are in good agreement.

The Si abundance was obtained from the Si\,I lines at 3905\,{\AA} and 4102\,{\AA}. Five Sc lines, six Ti\,I lines and 25 Ti\,II lines were employed to derive the respective abundances with good precision. Ti\,I and II agree within $0.08$\,dex. Three Cr\,I lines, three weak Mn\,I lines, 15 Co\,I lines, and 14 Ni\,I lines were also detected. Finally, we estimated a surprisingly low upper limit for the Zn abundance from the Zn\,{I} line at 4810\,{\AA} of $\mbox{[Zn/Fe]}<0.23$. This places J1808$-$5104 at the very bottom end of halo star Zn abundance range (see Figure~\ref{fig:abundplot}) at its [Fe/H] abundance.

Abundances of neutron-capture elements were determined for Sr and Ba from two Sr\,II lines and one weak Ba\,II line at 4554\,{\AA}. They are both significantly below solar ratios, very similar to values found for other halo and dwarf galaxy stars at these metallicities. Upper limits for Y, Zr and Eu were obtained from the lines at 4900\,{\AA}, 4149\,{\AA} and 4129\,{\AA}, respectively. The Y upper limit of $\mbox{[Y/Fe]}<-0.07$ indicates a low abundances; the Zr and Eu upper limits are too high to be very meaningful. However, even mild $r$-process enhancement can be excluded given the low Sr and Ba abundances. 

Standard deviations of individual line measurements for each element are taken as random abundance uncertainties. 
We assign a nominal minimum uncertainty value of 0.05\,dex for all species that have a standard deviations of $<0.05$\,dex. Further, we assign a nominal minimum uncertainty of 0.1\,dex in two cases: (i) abundances of elements with just one line measured; and (ii) standard deviations that are less than 0.1\,dex and the number of measured lines is three or less. The uncertainties are given in Table~\ref{Tab:abund}. Systematic uncertainties can be assessed from varying each stellar parameter by its uncertainty and re-determining the abundances. Typical final uncertainties ($\sigma_{rand} + \sigma_{sys}$) are about 0.15-0.25\,dex.

\begin{table}
\centering
\caption{Magellan/MIKE Chemical Abundances of J1808$-$5104.}
\label{Tab:abund}
\begin{tabular}{lrrrrr}
\hline
Species& $N$ & $\log\epsilon (\mbox{X})$ & $\sigma$&  [X/H]& [X/Fe] \\
       &     &       (dex)               &         & (dex) & (dex) \\
\hline
Li\,I (syn)  &  1   &  +1.38   &  0.10   &\ldots     &  \ldots\\
C (syn)      &  2   &  +4.91   &  0.20   &  $-$3.52   &  0.38\\
N (syn)      &  1   &$<+$3.47  &\ldots   &  $<-$4.36  &  $<-$0.20\\
O (syn)      &  2   &  +6.10:  &  0.20   &  $-$2.59:   &  1.25:\\
Na\,I        &  2   &  +2.08   &  0.10   &  $-$4.16   &  $-$0.32\\
Mg\,I        &  7   &  +4.09   &  0.05   &  $-$3.51   &  0.34\\
Al\,I        &  2   &  +1.60   &  0.10   &  $-$4.85   &  $-$1.01\\
Si\,I        &  2   &  +3.81   &  0.10   &  $-$3.70   &  0.14\\
Ca\,I        &  7   &  +2.49   &  0.05   &  $-$3.85   &  0.00\\
Sc\,I        &  5   &$-$1.02   &  0.05   &  $-$4.17   &  $-$0.32\\
Ti\,I        &  6   &  +1.20   &  0.05   &  $-$3.75   &  0.09\\
T\,II        &  25  &  +1.12   &  0.05   &  $-$3.83   &  0.02\\
Cr\,I        &  3   &  +1.08   &  0.10   &  $-$4.56   &  $-$0.71\\
Cr\,II       &  1   &  +1.89   &  0.10   &  $-$4.01   &  0.09\\
Mn\,I        &  3   &  +0.26   &  0.10   &  $-$5.17   &  $-$1.33\\
Fe\,I        &  96  &  +3.66   &  0.05   &  $-$3.84   &  0.00\\
Fe\,II       &  4   &  +3.55   &  0.05   &  $-$3.95   &  $-$0.11\\
Co\,I        &  15  &  +1.33   &  0.05   &  $-$3.66   &  0.18\\
Ni\,I        &  14  &  +2.12   &  0.05   &  $-$4.10   &  $-$0.26\\
Zn\,I        &  1   &$<+$0.94  &\ldots   & $<-$3.62   &  $<$0.23\\
Sr\,II (syn) &  2   &$-$1.85   &  0.10   &  $-$4.72   &  $-$0.87\\
Y\,II (syn)  &  1   &$<-$1.75  &\ldots   &  $<-$3.96  &  $<-$0.07\\
Zr\,II (syn) &  1   &$<-$0.60  &\ldots   &  $<-$3.18  &  $<+$0.66\\
Ba\,II (syn) &  1   & $-$2.36  &  0.10   &  $-$4.54   &  $-$0.70\\
Eu\,II (syn) &  1   &$<-$2.20  &\ldots   &  $<-$2.72  &  $<+$1.12\\
\hline
\end{tabular}
\end{table}

\section{Constraints on the progenitor star of J1808$-$5104 and its birth environment}\label{sec:discussion}

The abundance signature of light elements produced in fusion processes observed in J1808$-$5104 generally agrees extremely well with established abundance trends by other metal-poor stars down to about ${\metal}\sim-4.2$. This agreement can be seen in Figure~\ref{fig:abundplot}. 
Overall, these abundance trends are thought to reflect chemical enrichment by early core-collapse supernova(e) that exploded prior to the births of J1808$-$5104 and other, similar metal-poor stars. Moreover, as can be seen from Figure~\ref{fig:abundplot}, the gas from which these objects formed must have been very well mixed to help erase any local abundance anomalies or potential variations of the supernova yields that would have enriched these birth gas clouds \citep{cayrel2004}.

Interestingly, the same trends are found for metal-poor stars in dwarf galaxies, both in the ultra-faint dwarfs and classical dwarfs \citep[e.g.,][]{cohen09,frebel10,gilmore13,Simon2019}. For those stars, we know that they formed in distinct places, namely their respective host galaxies. For halo stars, we do not know their origins, but considering hierarchical assembly and the formation process of the Galactic halo, as well as their chemically primitive nature, we can assume that these stars also formed in accreted systems a long time ago. This would imply that all these metal-poor stars represent a large number of birth places of which each was chemically enriched by local supernovae. Yet, all these stars show near-identical abundance patterns. This points to a universal enrichment history, well mixed gas with mixing driven by robust processes, and/or similar supernova yields across all these environments \citep{frebel10, chan17}. 

There are several exceptions to these well-behaved abundance trends of light elements. Na shows much more scatter than other elements but reasons for that remain unclear. Carbon is known to display extreme variations, particularly at the lowest metallicities \citep{yoon16}. Nitrogen also varies significantly. About 60\% of metal-poor stars with ${\metal}\lesssim -3.5$ exhibit strong overabundances of carbon of ${\metal}>0.7$ \citep{placco14}. The carbon enhanced stars have been suggested to point to a specific birth environment, e.g. to minihalos \citep{cooke14} enriched by individual faint supernovae undergoing a mixing and fallback episode that would lead to large carbon and low iron yields \citep{UmedaNomotoNature}. On the contrary, metal-poor stars with abundances closer to the solar ratio, such as J1808$-$5104, more likely formed from well-mixed gas, perhaps in somewhat larger systems that hosted more than one progenitor supernova. The low N abundance would principally support this scenario also. Similarly, most stars in dwarf galaxies are not carbon enhanced. In particular, the larger classical dwarf galaxies show little evidence for a significant population of carbon rich stars \citep{tafelmeyer10, simon15,Chiti2020} perhaps because they formed from larger building blocks or accreted gas quickly to grow to large enough for efficient mixing to take effect. The ultra-faint dwarfs contain a small fraction of carbon-rich stars (e.g., \citealt{norris10_seg}) but contain primarily stars with abundances signatures very similar to that of J1808$-$5104 \citep{norris10booseg}. These systems might have already formed from multiple building blocks, which could represent a mix of progenitor systems that produced some (or even no) carbon enhanced stars.

Assuming then that J1808$-$5104 formed in one of the earliest systems in the early universe, the progenitor supernova must thus have either not undergone a mixing and fallback mechanism, or there were not enough fallback supernovae to dominate the resulting chemical composition of the gas over the enrichment by ordinary core-collapse supernovae. After all, metal mixing processes were likely very efficient following the energy injection by supernovae and subsequent recovery time of the system, before forming the next generation of stars \citep{greif10}. 

\begin{figure*}
\includegraphics[width=1.05\textwidth]{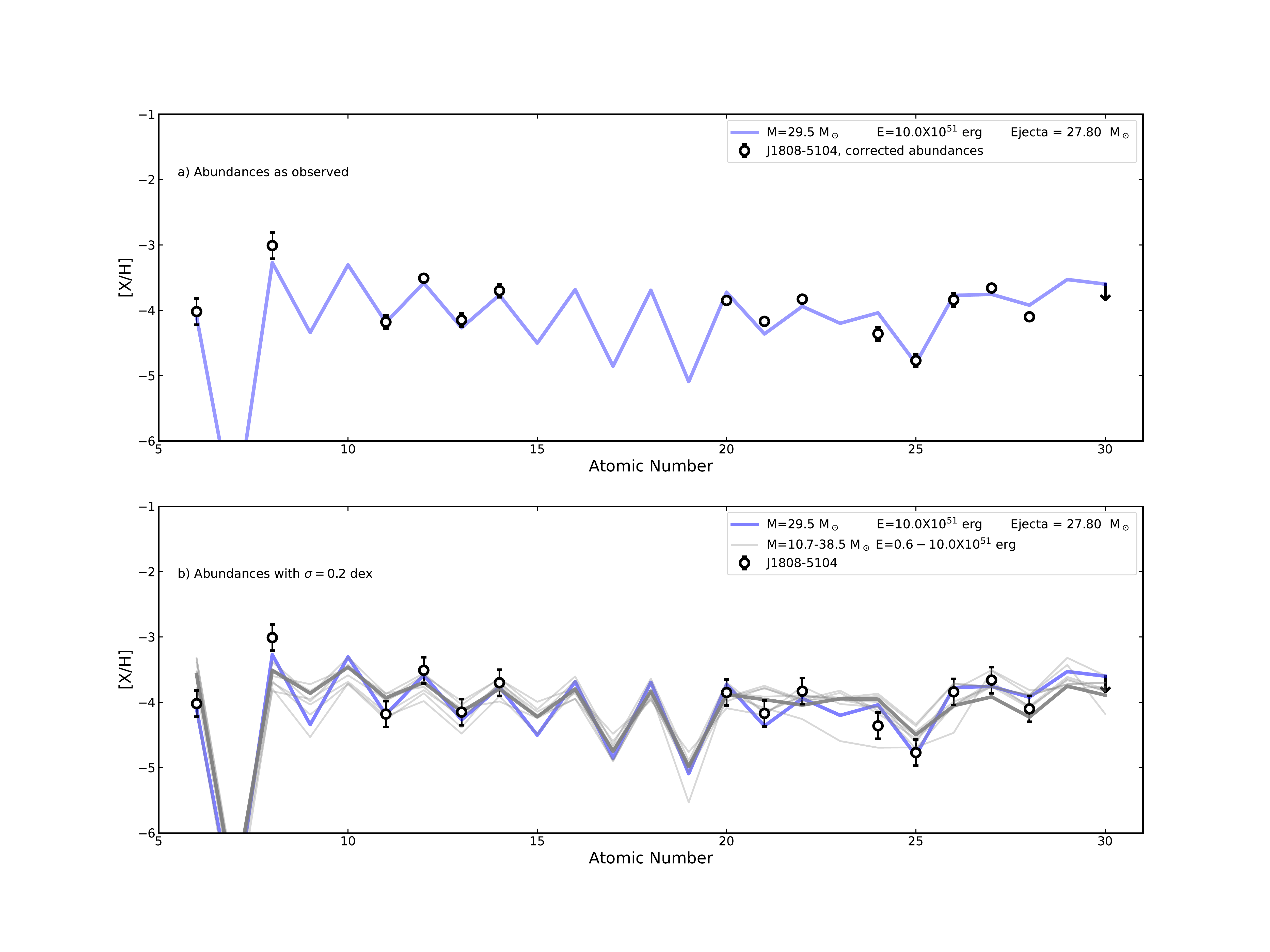} 
\caption{Best fit models to the chemical abundances of J1808$-$5104. The open circles represent abundance reported in Table~\ref{Tab:abund}. Error bars represent our adopted uncertainties in Table~\ref{Tab:abund} (top panel), and 0.2\,dex (bottom panel). Lines depict the nucleosynthetic yield of the best fit models \citep{heger10}, with the blue line showing the best fit to the data. This model is also the best fit in 62\% of the $10^4$ realizations. Masses and explosion energies are listed in the legend at the upper right of each panel. Gray lines depict additional models with a variety of masses and energies that have been found to be best fits in a fraction of the $10^4$ realization, for each uncertainty case.\label{sn_fit}}
\end{figure*}

To test this idea, we attempted to model the light element abundance signature of J1808$-$5104 with theoretical Pop\,III supernova nucleosynthesis yields from \citet{heger10}. Employing their $\chi^2$ matching algorithm\footnote{\url{http://starfit.org}} provides insights into the putative progenitor(s) (e.g., stellar mass and supernova explosion energy) of metal-poor stars with {\metal} $<-3.0$.
This approach has been applied to a variety of metal-poor stars in the literature \citep[e..g.,][]{placco15b,placco16b,placco16,roederer2016,Placco2020}.
We note that the nucleosynthesis models are all (S4) fallback models with masses from 10 to 100\,M$_\odot$, and explosion energies from $0.3 \times 10^{51}$\,erg to $10 \times 10^{51}$\,erg. However, the mass ejected during the explosion is also given and not all models have significant amounts of fallback.

Running the {\sc{starfit}} algorithm for J1808$-$5104 using the chemical abundances from Table~\ref{Tab:abund}, suggest a series best-fit models (all with $\chi^{2} \sim 6$) with a progenitor stellar mass of M~=~29.5\,M$_\odot$, high explosion energy of E~=~$10.0\times10^{51}$\,erg and little mixing with a range of $\log (f_{mix}) \sim-3.0$ to $-0.6$. With the estimated ejecta of 27.8\,M$_\odot$, essentially no fallback appears to occur. We note that we used a 3D-corrected carbon abundance of $\mbox{[C/H]}=-4.02$, based on the \citep{gallagher16} estimate made in Section~\ref{sec:chem}. For comparison, \citet{Spite2019} suggested a $-0.4~\pm~0.1$\,dex correction. Similarly, for O, we utilized a 3D-corrected value. Since our abundance is uncertain, we opted adopt the correction $\mbox{[O/H]}=-3.08$, which includes a $-$0.6\,dex correction of the [O/H] reported in \citet{Spite2019}, for our fitting procedure. Other abundances were also corrected before fitting the abundance pattern. We applied a Na correction of $-$0.04\,dex, using our measured EW and the online calculator {\sc{INSPECT}}\footnote{\url{http://www.inspect-stars.com}} \citep{Lind_nlte_Na}, an Al correction of +0.7\,dex \citep{Nordlander2017_Al_NLTE,Roederer2021_Al_NLTE}, a Cr correction of 0.2\,dex \citep{Bergemann2010_Cr_NLTE,Cowan2020_cr_NLTE}, and Mn correction of 0.4\,dex \citep{Bergemann2008_Mn_NLTE,Sneden2016_Mn_NLTE}. Finally, Zn generally constrains the explosion energy of the progenitor but that we only have an upper limit. Given the low upper limit, and considering the bulk of halo stars at $\mbox{[Fe/H]}\sim-3.5$, it could be argued that the true Zn abundance is not much below the numerical value of our upper limit determination. As such, for the fitting purposes, we treat our upper limit as a measured abundance albeit with a larger error bar of 0.3\,dex. The top panel in Figure~\ref{sn_fit} shows our abundances overlaid with the nucleosynthesis yields of the 29.5\,M$_\odot$ and $10.0\times10^{51}$\,erg model.



Next, we statistically checked on the robustness of the fitting results by generating 10,000 abundance patterns for J1808$-$5104, by re-sampling the $\log\epsilon (\mbox{X})$ values from Table~\ref{Tab:abund} with fixed uncertainties of $\sigma = 0.2$\,dex for all species except for those that have larger uncertainties already (i.e., C, O, Zn). Running the {\sc{starfit}} code for each re-sampled pattern (and determining its respective best-fit model) results in 20 best-fitting models with a range of parameters. For simplicity, we chose to ignore any best fit models if only 1-10 realizations favored those. Overall, 62\% of the 10,000 patterns are matched best by the model with 29.5\,M$_\odot$ and $10.0\times10^{51}$\,erg (i.e. the same no-fallback model as found for the original abundance pattern above). The remaining realizations ($\sim 38\%$) are best fit by a variety of different models, with stellar masses ranging from M= 10.2 to 38\,M$_\odot$ and explosion energies of E~=0.6 to $10.0\times10^{51}$\,erg. 
The results are shown in the bottom panel of Figure~\ref{sn_fit}, where we show the abundances with the $\sigma = 0.2$\,dex uncertainties overlaid with the 20 best fitting models, many of which yield very similar patterns.

We conclude that J1808$-$5104 most likely formed in an environment that experienced the enrichment by a massive Population\,III hypernova with a high explosion energy and little to no fallback. This is a similar result as what has been found from many the analysis of other other similar metal-poor stars, adding to the body of evidence that the first stars were predominantly massive in nature.

This origin scenario is supported by two additional lines of evidence. First, the Sr and Ba abundances of J1808$-$5104 are extremely low, $\mbox{[Sr/H]}\sim-4.7$ and $\mbox{[Ba/H]}\sim-4.5$. These low values are typical for stars found in ultra-faint dwarf galaxies and also some of the classical dwarfs. Furthermore, the star has $\mbox{[Sr/Ba]}= -0.17$. This value is not far removed from what is typical for the (main) $r$-process, $\mbox{[Sr/Ba]}\sim-0.4$. A limited $r$-process, with $\mbox{[Sr/Ba]}>0.5$ \citep{Frebel2018}, can clearly be ruled out, so can the $s$-process ($\mbox{[Sr/Ba]}<-1$).

\begin{figure}
\begin{center}
\includegraphics[clip=true,width=0.5\textwidth]{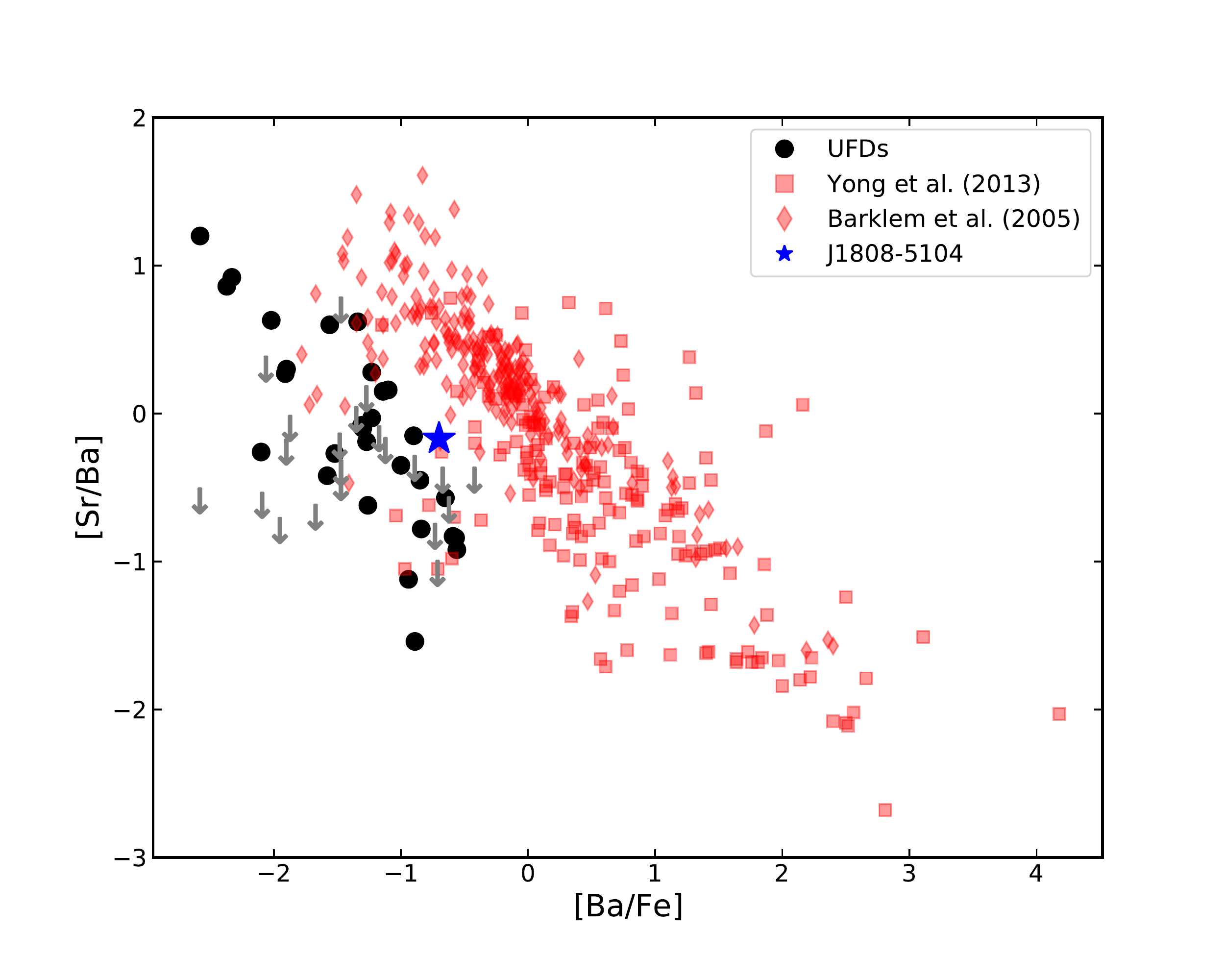} 
\caption{[Sr/Ba] as a function of [Ba/Fe] for ultra-faint dwarf galaxy stars and J1808$-$5104 (black points represent measurements, downward arrows represent upper limits) and other halo metal-poor stars (red squares) adopted from \citep{Yong2013}. Blue asterisk represents J1808$-$5104 abundances. \label{fig:sr_ba}}
\end{center}
\end{figure}

Figure~\ref{fig:sr_ba} shows [Sr/Ba] as a function of [Ba/Fe] for ultra-faint dwarf galaxies stars overplotted with halo metal-poor stars (red squares and diamons) adopted from \citet{Yong2013} and \citet{barklem05}, respectively\footnote{Literature data collection of ultra-faint dwarf galaxy stars taken from  \url{https://github.com/alexji/alexmods}}. Black points represent ultra-faint dwarf galaxy stars with Sr and Ba measurement, downward arrows represent stars with upper limits on Sr and/or Ba abundances. Plotting [Sr/Ba] against [Ba/Fe], J1808$-$5104 is found below the main trend set by metal-poor halo stars, in a region that is characteristically populated by stars in the dwarf galaxies. This all suggests J1808$-$5104's neutron-capture elements to possibly be provided by one supernova or explosive event only. Following arguments laid out in \citet{ji16b}, a level of $\mbox{[Ba/H]}\sim-5$ is reached by the yields of one supernova if the gas mass into which the yield is diluted into is 10$^6$\,M$_\odot$. This adds confidence to the scenario that the star formed in a sparse system with only one or few SNe progenitors. This is also already indicated by the low [Fe/H] of the star, which is suggestive of a 10$^6$\,M$_\odot$ birth cloud (and assuming a canonical Fe yield of 0.1\,M$_\odot$). It is thus indicative that stars with $-4.5\lesssim{\metal}\lesssim-4.0$ and
$0\lesssim$~[C/Fe]~$\lesssim0.7$ all formed in similar
environments that already experienced some degree of chemical
homogeneity, while also showing clear signs in the form of low neutron-capture abundances of only one, or at most a small number of progenitor stars.


\subsection{Kinematic Signature}

\begin{figure*}
\begin{center}
\includegraphics[clip=true,width=\textwidth]{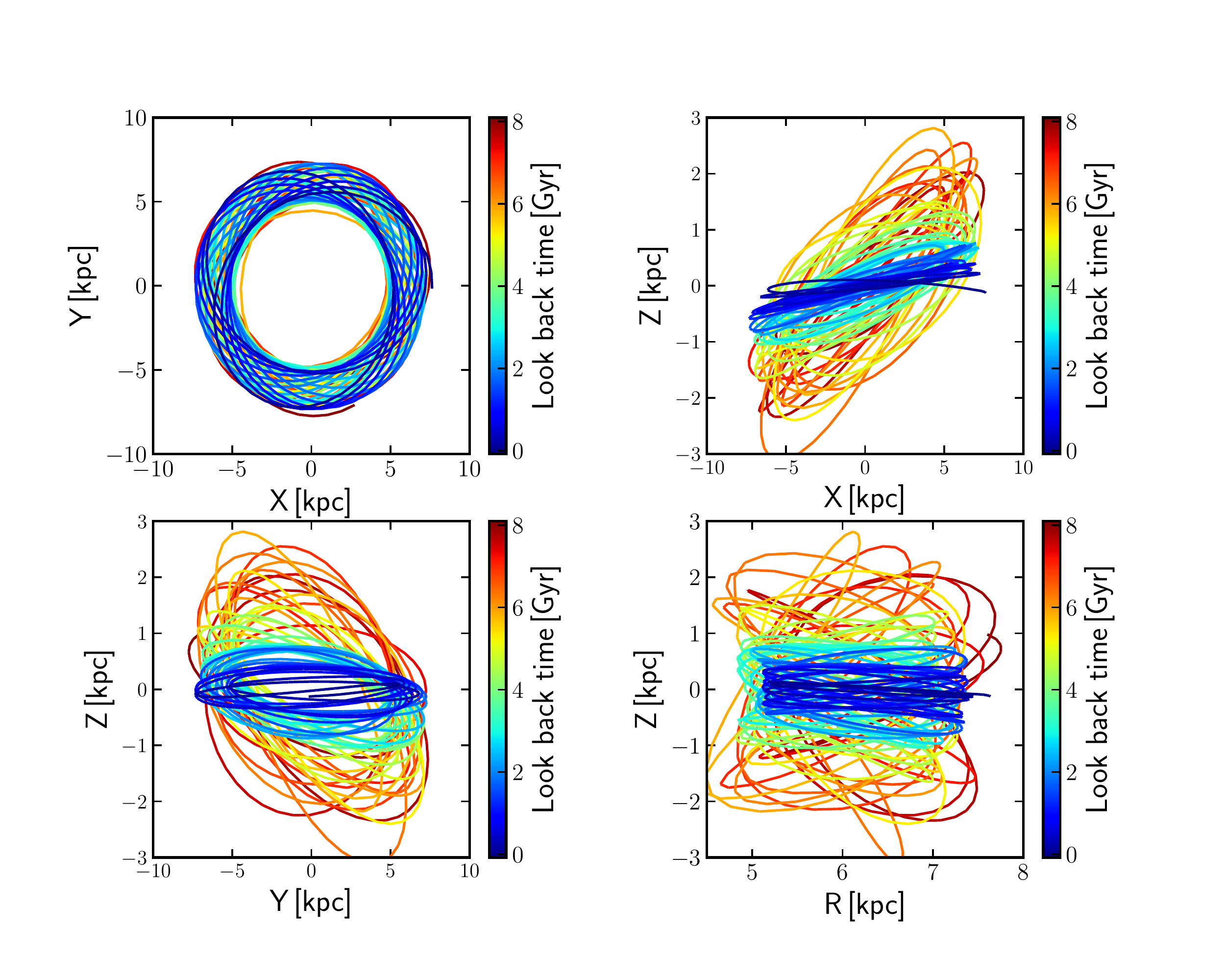}
\caption{Projections of the J1808$-$5104 orbit in the X-Y (upper left panel), X-Z (upper right panel), Y-Z (lower left panel), and R-Z (lower right panel). The X-Y plan suggests that J1808$-$5104 has very circular orbit. The R-Z plan suggests that J1808$-$5104 can reach the solar neighborhood R~$\sim 8$\,kpc. This orbital integration was performed using the \texttt{ORIENT} potential~$\#483868$. \label{fig:orient_orbits}}
\end{center}
\end{figure*}

Investigating the long-term orbital history of J1808$-$5104 can add a new dimension to comprehensively probe its origin. Detailed space motion for J1808$-$5104 can be derived by combining astrometric information obtained by Gaia DR3 \citep{Gaia_DR3} and the systemic radial velocity (RV = 14.8 \,km\,s$^{-1}$ at phase = 0; see Figure~\ref{orbital_period}). To perform this investigation in a statistical way, we generate 10,000 realizations of the celestial positions ($\alpha$, $\delta$), proper motions ($\mu_{\alpha}$, $\mu_{\delta}$), and the systemic RV value using a normal distribution and associated uncertainties. 

We then assume that the Sun is located at R$_\odot =8.178 \pm 0.013$\,kpc from the Galactic center \citep{Gravity_Collaboration2019}, $z_{\odot}= 20.8 \pm 0.3$\,pc above the Galactic plane, and has peculiar motion $U_{\odot} =11.1 \pm 0.72$\,km\,s$^{-1}$ \citep{Bennett2019}, $V_{\odot}= 12.24 \pm 0.47$\,km\,s$^{-1}$, and $W_{\odot}= 7.25 \pm 0.36$\,km\,s$^{-1}$ \citep{Schonrich2010}. We take V$_{LSR}$ = $220$ km\,s$^{-1}$ \citep{Kerr1986}. For each realization, we calculate Galactocentric coordinates ($X,Y,Z$), rectangular Galactic ($U,V,W$), and cylindrical Galactocentric coordinates ($V_{R}, V_{\phi}, V_{z}$), as described in \citet{Mardini2022}. We also compute orbital parameters (Z$_{max}$, r$_{apo}$, r$_{peri}$, eccentricity), for the past 8\,Gyr using our time-varying galactic potentials, \texttt{ORIENT}\footnote{\url{https://github.com/Mohammad-Mardini/The-ORIENT}} \citep[for more details, we refer the readers to ][]{Mardini_2020}.

Figure~\ref{fig:orient_orbits} shows the projections of the long term orbital evolution of J1808$-$5104 in various planes, for one of the 10,000 realizations, using the \texttt{ORIENT} potential~$\#483868$. The X-Y plane suggests that J1808$-$5104 is on a quasi-circular orbit (e~=~$0.22 \pm 0.01$). Furthermore, the R-$|Z|$ plane suggests that J1808$-$5104 currently (cycles where the colour tends toward blue, that is short look back time) resides in the Galactic thin disk, orbiting with a radius of $\sim 8$\,kpc from the Galactic center. While at a long look back time, the star might travel 3\,kpc above and below the plane of the present-day Galactic disk (cycles where the colour tends toward red).

The Galactic model used to evolve the orbit has time-varying potential derived from best-fitting a subhalo in a large scale cosmological simulation that is similar in some metrics to the present-day Milky-Way \citep{Mardini_2020}. While the model is a composition of an NFW sphere and a Miyamoto-Nagai disk at every given time, the seven free parameters are time-varying. That includes the orientation of the disk, which has two free angles and is oriented such that it coincides with the X-Y plane at the present-day. The particular \texttt{ORIENT} model has a disk that is inclined by up to $22^\circ$ relative to present day, this largest inclination occurs at a look back time of 6\,Gyr or at redshift of $z \sim 0.6$. The large $Z$-values seen in Figure~\ref{fig:orient_orbits} around that time therefore reflect mostly the inclination of the disk itself: even if the star were on a perfectly planar orbit, its $Z$ ordinate would reach values as large as 3\,kpc (as this is the approximate Galactocentric distance of 8\,kpc times the sine of the maximum inclination angle)\footnote{A right triangle with one point at the centre of the Galaxy, another point at the star at its apocentre, and the third point in the present-day Galactic plane under the star. In this convention, the r$_{apo}$ and Z$_{max}$ are the sides of the triangle across from the disk's inclination ($\alpha$), so $\sin (\alpha)$ = Z$_{max}$/r$_{apo}$ by definition.}.

\begin{figure}
\begin{center}
\includegraphics[clip=true,width=0.5\textwidth]{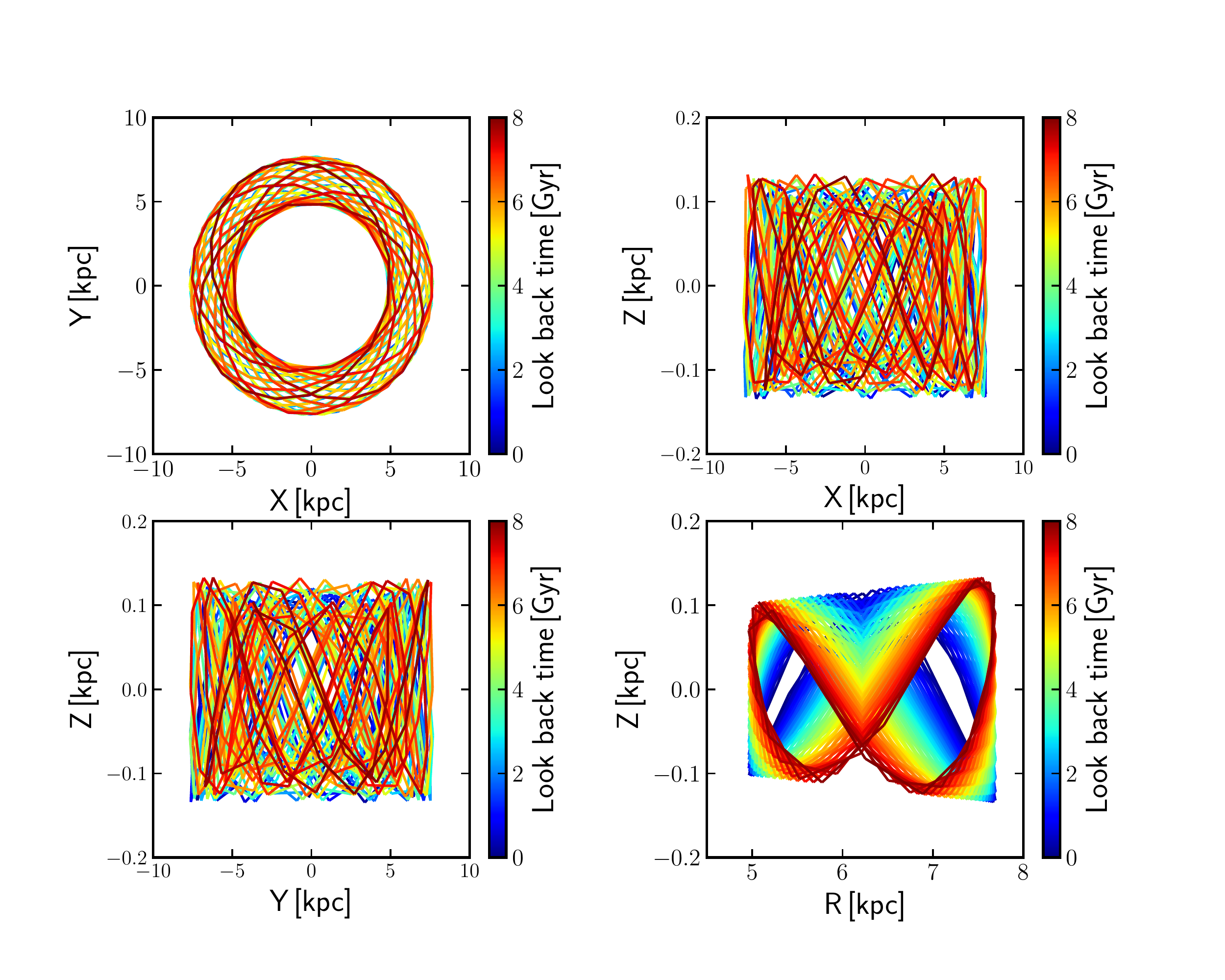}
\caption{Same as in Figure~\ref{fig:orient_orbits}, but using \texttt{galpy} and \texttt{MWpotential2014} \citep{Bovy2015}. \label{fig:orbits}}
\end{center}
\end{figure}

In addition, and for comparison purposes, we performed another backward orbital integration for J1808$-$5104 using \texttt{galpy}\footnote{\url{https://docs.galpy.org/en/v1.8.0/}} and its invariable Galactic potential \texttt{MWpotential2014} \citep{Bovy2015}. 
It is important to recall here that the size and the mass of the Milky Way components in \texttt{MWpotential2014} are assumed to remain constant with time. Using such an idealized potential would not mimic the realistic formation and evolution history of the Milky Way which itself was built from smaller accreted satellites \citep[e.g.,][]{hierarchical1,hierarchical2,hierarchical3}. Figure~\ref{fig:orbits} shows the same projections as in Figure~\ref{fig:orient_orbits}. Apparently, the static orbits do not change significantly with redshift and the phase-space coordinates of J1808$-$5104 at high redshifts do not extend to higher Galactocentric distances either. 

The main difference between Figures~\ref{fig:orient_orbits} and \ref{fig:orbits}, or the integration in time-varying and time-static potential respectively, is that no conserved quantities exist in the former case. In addition to the variation in the model's disk orientation mentioned above, other quantities vary as well, some quite significantly over the integration period. For example, the disk's mass increases at approximately constant rate from $\sim 4\times 10^{10} \,\mathrm{M}_\odot$ at a look back time of 8 Gyr to $\sim 7\times 10^{10} \,\mathrm{M}_\odot$ at a look back time of 1.5 Gyr (after which it does not vary significantly). Orbits integrated with \texttt{ORIENT} models may therefore demonstrate irregular behaviour compared to those integrated with \texttt{galpy}, depending on the model's specific cosmic history.

Finally, we probe the thin disk membership of J1808$-$5104 using the diagnostic tool developed in \citet{Mardini2022}, based on stellar actions and velocities, to qualitatively assign individual stars to one of the traditional Galactic components. All of the J1808$-$5104 10,000 realizations suggest that the star to be well confined to the thin disk.

To learn about the overall origin scenario of J1808$-$5104, two possible pathways can be considered that are based on the fact that the star has been located in the thin disk for billions of years. (1) J1808$-$5104 could have originated in a satellite galaxy that was accreted into the disk and fully disrupted at early times in the formation history of the Milky Way. (2) J1808$-$5104 could have been one of the earliest in-situ stellar births as part of the formation of the primordial thin disk which came together from small building blocks. 
The fact that the star has a very old age of $\sim13.5$\,Gyr \citep{Schlaufman2018} needs to be taken into account when addressing these scenarios.

Regarding scenario (1), some theoretical models of the formation of the Galactic disk predict an old thin disk population to be built up from satellite(s) debris \citep[for example, see figure~8 in][]{Abadi2003}. However, the contribution of the satellite heavily depends on its orbit and the level of the dynamical friction \citep{Statler1988}. In this picture, the core of the satellite should be dense enough to survive tidal disruption up to the time that it circularized its orbit within the disk (i.e. interacts strongly with the disk and deposit significant fraction of its stars). The still only low fraction of the discovered metal-poor ({\metal} $<-3$) stars with thin disk-like kinematics \citep[e.g.,][]{Bensby2014,Sestito2019,Carter2020,Cordoni+2020,Matteo2020,Venn2020, Mardini2022} furthermore suggests that this scenario likely does not explain the origin of J1808$-$5104.

\begin{figure}
\begin{center}
\includegraphics[clip=true,width=0.5\textwidth]{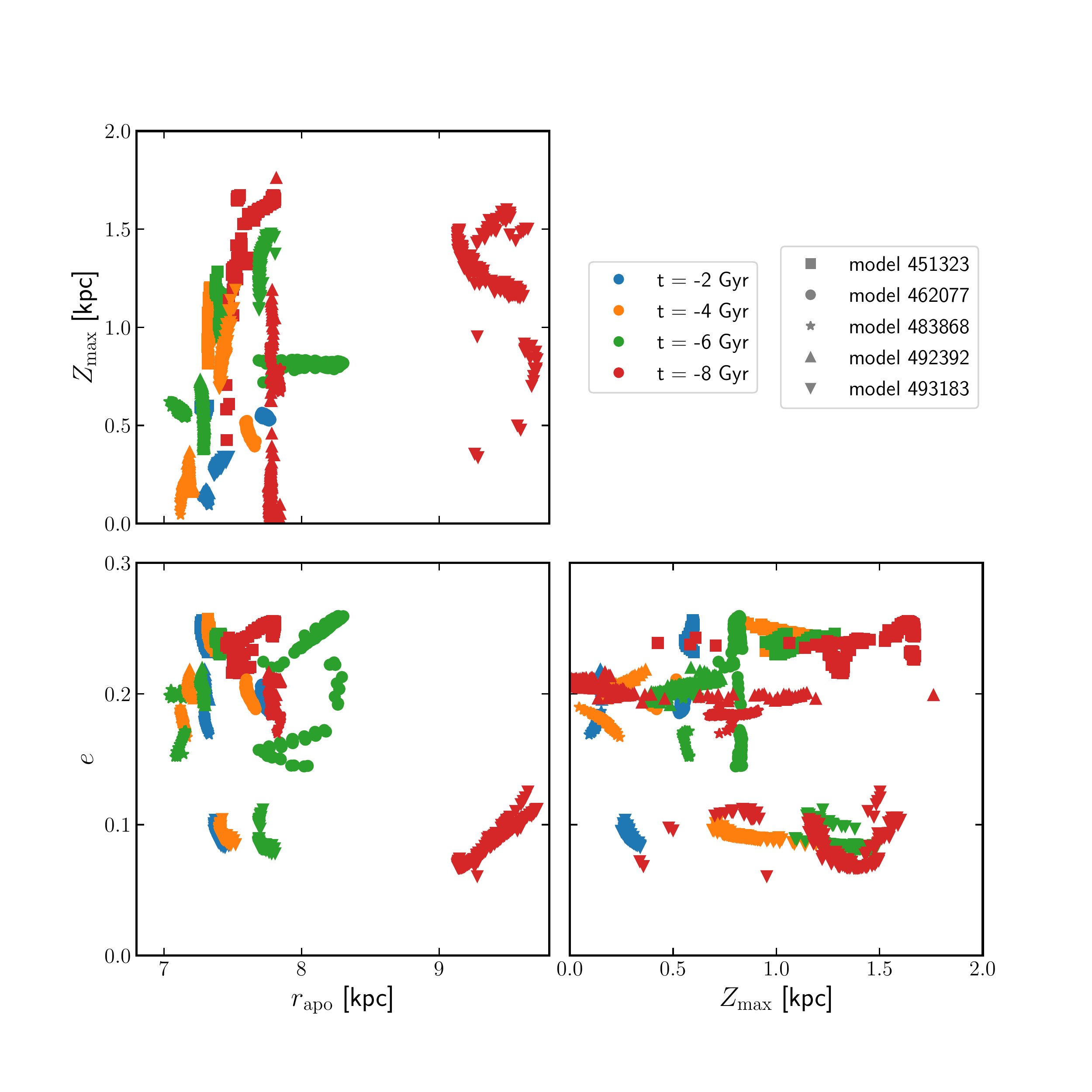}
\caption{Distributions of e, r$_{apo}$, and Z$_{max}$ for 200 randomly selected realizations at four different cosmic times. Different symbols represent different \texttt{ORIENT} models. Different colors represent different cosmic time. All of these data points maintain thin disk kinematics over the past 8\,Gyr. \label{fig:orient_test}}
\end{center}
\end{figure}

The second scenario then requires the absence of dynamical interactions with the spiral arms and/or merging satellite(s) to not heat up the orbit of J1808$-$5104 (i.e. increase e and Z as a function of time). We investigated this scenario by tracing back the orbital history of each one of the 10,000 realizations obtained with \texttt{ORIENT}, to check whether J1808$-$5104 can maintain its thin disk kinematics during the past 8\,Gyr. Figure~\ref{fig:orient_test} shows e, r$_{apo}$, and Z$_{max}$ for 200 realizations (each data point represents one realization)\footnote{Our version containing results of the entire 10,000 realizations is very crowed. Therefore, we show the results for 200 randomly selected realizations.}, of five different \texttt{ORIENT} potentials (each symbol represents one model), at four different cosmic times (each color represents one cosmic time). Notice that Z$_{max}$ is calculated with respect to the model's disk orientation at a given cosmic time. All of these data points holds thin disk kinematics. Thus, it appears physically quite possible for J1808$-$5104 to survive these dynamical interactions and maintain its thin disk kinematics over billions of years. 
 

In summary, our chemical abundances and kinematics results suggest, paired with the old age of the system \citep[$\approx 13$\,Gyr][]{Schlaufman2018}, that J1808$-$5104 is the most primitive thin disk star known. It likely formed at the earliest epoch of the hierarchical assembly of the Milky Way, and would thus be an ancient member of the primordial thin disk.

\section{Conclusions}\label{sec:Conclusions}
In this paper, we present a comprehensive chemo-dynamical analysis of the most metal-poor thin disk star 2MASS~J18082002$-$5104378. We provide further five radial-velocity (RV) measurements based on Magellan/MIKE high resolution spectra. These RV measurements suggest that J1808$-$5104 is in a binary system. The system has an orbital period of $P = 34.7385_{-0.2}^{+0.2}$\,days and line-of-sight velocity of RV = 14.8 \,km\,s$^{-1}$. 

We report on the first detection of the Ba\,{II} line at 4554\,\AA\ for J1808$-$5104. The observed chemical pattern suggest that J1808$-$5104 exhibits mild enhancements in the $\alpha$-elements, and no enhancements in either carbon ($\mbox{[C/Fe]} = 0.38$ $\pm 0.10$) or neutron-capture elements ([Sr/Fe]= $-0.87$ $\pm 0.10$ and [Ba/Fe]= $-0.70$ $\pm 0.10$); indicating that J1808$-$5104 was formed in chemically primitive cloud that experienced relatively few enrichment events. We compare the light elements abundance pattern to theoretical yields of Population\,III adopted from \cite{heger_woosley10}. The best fit model suggest a progenitor with stellar mass of 29.5\,M$_\odot$ and explosion energies 10$\times 10^{51}$\,erg. In general, the comparison suggest a fallback with no mixing supernovae to be responsible for the chemical enrichment of J1808$-$5104.

We also perform a comprehensive study of the possible orbital evolution of 10,000 J1808$-$5104-like stars using our time-dependent galactic potential the \texttt{ORIENT} and the diagnostic tool developed in \citet{Mardini2022}. The results show that all of the J1808$-$5104-like stars maintain a quasi-circular orbits, Z$_{max} < 1$ and still bound to the Galaxy. In general, these orbits exclude the possibility that J1808$-$5104 has an accretion origin, and suggest it being member of the primordial thin disk.

\section*{Acknowledgements}
This work is supported by Basic Research Grant (Super AI) of Institute for AI and Beyond of the University of Tokyo. M.k.M. acknowledges partial support from NSF grant OISE 1927130 (International Research Network for Nuclear Astrophysics/IReNA). A.F. acknowledges support from  NSF CAREER grant AST-1255160 and NSF grant AST-1716251. A.C. is supported by a Brinson Prize Fellowship at the University of Chicago/KICP. The work of V.M.P. is supported by NOIRLab, which is managed by the Association of Universities for Research in Astronomy (AURA) under a cooperative agreement with the National Science Foundation. I.U.R.\ acknowledges support from NSF grant AST~1815403/1815767 and the NASA Astrophysics Data Analysis Program, grant 80NSSC21K0627. This work made use of the NASA's Astrophysics Data System Bibliographic Services.

This work has made use of data from the European Space Agency (ESA) mission
{\it Gaia} (\url{https://www.cosmos.esa.int/gaia}), processed by the {\it Gaia}
Data Processing and Analysis Consortium (DPAC,
\url{https://www.cosmos.esa.int/web/gaia/dpac/consortium}). Funding for the DPAC
has been provided by national institutions, in particular the institutions
participating in the {\it Gaia} Multilateral Agreement.



\bibliographystyle{mnras}
\bibliography{references}

\bsp	
\label{lastpage}
\end{document}